\documentclass[a4paper, accepted=2023-03-16]{quantumarticle}

\pdfoutput=1
\usepackage[numbers,sort&compress]{natbib}
\usepackage[english]{babel}
\usepackage{letltxmacro}
\usepackage{latexsym}
\LetLtxMacro{\ORIGselectlanguage}{\selectlanguage}
\makeatletter
\DeclareRobustCommand{\selectlanguage}[1]{%
  \@ifundefined{alias@\string#1}
    {\ORIGselectlanguage{#1}}
    {\begingroup\edef\x{\endgroup
       \noexpand\ORIGselectlanguage{\@nameuse{alias@#1}}}\x}%
}
\newcommand{\definelanguagealias}[2]{%
  \@namedef{alias@#1}{#2}%
}
\makeatother
\definelanguagealias{en}{english}
\definelanguagealias{English}{english}
\usepackage[T1]{fontenc}
\usepackage{array}
\usepackage{graphicx}
\usepackage{amsmath}
\usepackage{amsfonts}
\usepackage{amsthm}
\usepackage{amssymb,bbding}
\usepackage{bm}
\usepackage[title]{appendix}
\usepackage{color}
\usepackage[percent]{overpic}
\usepackage{soul} 
\usepackage{wasysym}
\usepackage{dsfont}
\usepackage{float}
\usepackage{braket}
\usepackage{cancel}
\usepackage{comment}
\usepackage{multirow}

\usepackage{enumitem}
\usepackage[dvipsnames]{xcolor}
\usepackage{graphicx}

\usepackage{mathtools}
\usepackage{comment}
\usepackage{relsize}
\usepackage{dsfont} 
\usepackage{commath}
\usepackage{titlesec}
\usepackage{makecell}

\usepackage{hyperref}
\hypersetup{
    pdffitwindow=false,         
    pdfstartview={FitH},        
    pdftitle={Characterization of variational quantum algorithms using free fermions}, 
    pdfsubject={Quantum physics journal article},
    pdfproducer={Gabriel Matos},
    pdfnewwindow=true,          
    colorlinks=true,            
    citecolor=blue,             
    filecolor=magenta,          
    urlcolor=blue,              
    bookmarksopen=true,
    bookmarksnumbered=true
}

\renewcommand*{\thesection}{%
  \arabic{section}\texorpdfstring{}{. }%
}
\renewcommand*{\thesubsection}{%
  \arabic{section}.\arabic{subsection}\texorpdfstring{}{. }%
}

\usepackage{cleveref}

\usepackage[utf8]{inputenc}
\usepackage[export]{adjustbox}

\setlist[itemize]{leftmargin=*}
\setlist[enumerate]{leftmargin=*}

\crefformat{equation}{Eq.~(#2#1#3)}
\Crefformat{equation}{Eq.~(#2#1#3)}
\crefrangeformat{equation}{Eqs.~(#3#1#4--#5#2#6)}
\crefname{figure}{Fig.}{Fig.}

\DeclareMathOperator*{\trace}{Tr}

\begin{document}

\title{Characterization of variational quantum algorithms using free fermions}

\author[1]{Gabriel Matos}
\email{pygdfm@leeds.ac.uk}
\orcid{0000-0002-3373-0128}
\author[2,3]{Chris N. Self} 
\orcid{0000-0002-1648-6725}
\author[1]{Zlatko Papi\'c} 
\orcid{0000-0002-8451-2235}
\author[4,5]{Konstantinos Meichanetzidis}
\author[6]{Henrik Dreyer}

\affiliation[1]{School of Physics and Astronomy, University of Leeds, Leeds LS2 9JT, United Kingdom}
\affiliation[2]{Quantinuum, Partnership House, Carlisle Place, London, SW1P 1BX, United Kingdom}
\affiliation[3]{Blackett Laboratory, Imperial College London, London SW7 2AZ, United Kingdom}
\affiliation[4]{Quantinuum, 17 Beaumont St., Oxford OX1 2NA, United Kingdom}
\affiliation[5]{Department of Computer Science, University of Oxford, Oxford OX1 3QD, United Kingdom}
\affiliation[6]{Quantinuum, Leopoldstrasse 180, 80804 Munich, Germany}

\theoremstyle{definition}
\newtheorem{theorem}{Theorem}
\newtheorem{lemma}[theorem]{Lemma}
\newtheorem{definition}[theorem]{Definition}

\date{March 11, 2023}

\begin{abstract}
We study variational quantum algorithms from the perspective of free fermions. By deriving the explicit structure of the associated Lie algebras, we show that the Quantum Approximate Optimization Algorithm (QAOA) on a one-dimensional lattice -- with and without decoupled angles -- is able to prepare all fermionic Gaussian states respecting the symmetries of the circuit. Leveraging these results, we numerically study the interplay between these symmetries and the locality of the target state, and find that an absence of symmetries makes nonlocal states easier to prepare. An efficient classical simulation of Gaussian states, with system sizes up to $80$ and deep circuits, is employed to study the behavior of the circuit when it is overparameterized. In this regime of optimization, we find that the number of iterations to converge to the solution scales linearly with system size. Moreover, we observe that the number of iterations to converge to the solution decreases exponentially with the depth of the circuit, until it saturates at a depth which is quadratic in system size. Finally, we conclude that the improvement in the optimization can be explained in terms of better local linear approximations provided by the gradients. 
\end{abstract}

\maketitle{}

\section{Introduction}

Variational quantum algorithms have recently received much attention as a potential candidate for demonstrating  quantum advantage. Originally proposed in the context of quantum chemistry~\cite{peruzzo_variational_2014} and classical optimization problems~\cite{farhi_qaoa_2014}, such algorithms have since found wide use as a tool for leveraging the current generation of noisy, intermediate scale quantum computers~\cite{preskill_nisq_2018} to tackle problems which are hard to solve classically. For that purpose they have been extended to a myriad of  domains, including machine learning, \cite{farhi_neural_2018, havlicek_supervised_2019, bokhan_cnn_2022, abbas_power_2021, schuld_qml_2019, cong_quantum_2019}, preparation of general condensed matter quantum states \cite{ho_hsieh_2019, xie_schwinger_2022, feulner_j1j2_2022, scala_witnessing_2022, schindler_sk_2022, warren_adaptive_2022, jattana_heisenberg_2022}, finance \cite{orus_finance_2019, egger_finance_2020}, molecular biology and biochemisty \cite{boulebnane_peptide_2022, outeiral_biology_2021}, and linear algebra
\cite{bravoprieto_linear_2019, chen_schmidt_2021, xiaosi_variational_2021}, and have been implemented in numerous quantum simulation platforms \cite{kandala_hardware_2017, grimsley_adaptive_2019, pagano_ising_2020, arute_hf_2020, weindfeller_scaling_2022}.

A major issue with this class of algorithms is that they are notoriously hard to optimize classically. Not only are they known to have an optimization landscape riddled with local minima \cite{huembli_loss_2021, lockwood_optimization_2022}, they also suffer from the phenomenon of ``barren plateaus", in which the gradients with respect to the optimization parameters vanish exponentially with system size \cite{wang_noise_induced_2021, mcclean_barren_2018}. Though several strategies have been proposed to deal with this problem \cite{li_vqe_2022, tao_laws_2022, kulshrestha_beinit_2022, zhang_gaussian_2022, liu_barren_2022, grimsley_avqe_2022, mele_transferability_2022} it remains an active area of research. 

Successfully employing parametrized circuits requires a balance between \emph{expressibility} and \emph{trainability}. Indeed, while universal circuit ansätze exist \cite{morales_universality_2020, biamonte_universal_2021}, problem-agnostic circuits typically present barren plateaus \cite{holmes_expressibility_2022}. Thus, it is crucial to design parameterized circuits that are able to adequately prepare or approximate the quantum state of interest, while not being overly expressive so as not to compromise its trainability. To this end, a characterization of the expressibility of these algorithms has been performed in several contexts, and strategies to systematically quantify it have been proposed \cite{du_efficient_2022, sukin_expressibility_2019, nakaji_expressibility_2021, abbas_power_2021, akshay_depth_2022}. A way to constrain the expressibility of a parameterized circuit is to employ \emph{problem-tailored ans\"atze}, such as the Hamiltonian Variational Ansatz \cite{wiersema_hva_2020}, among other strategies such as exploiting symmetries in the problem \cite{gard_efficient_2020, volkoff_gradients_2021, tsuchimochi_symmetry_2022, ender_modular_2022, lyu_symmetry_2022} or removing redundant parameters \cite{funcke_dimensional_2021, funcke_dimensional_2021_2}. 

Other factors not immediately linked to the expressibility of the circuit are known to affect the optimization, such as the boundary conditions used; these have been observed to greatly affect the success of optimization, and strategies have recently been proposed to address this issue \cite{sun_performance_2022}. Another set of such factors is related to the individual characteristics of states to be prepared, such as locality of correlations and entanglement
\cite{ho_ultrafast_2019, marrero_entanglement_2020, pavel_correlation_2021, uvarov_locality_2021, nikolay_vqe_2021, deller_interactions_2022}. Despite the extensive work done on this subject, there is no systematic characterization or theory explaining the interplay between all these aspects and their connection to expressibility.

Finally, it has been observed that an increase in the number of parameters in the circuit without a corresponding increase in the expressibility induces an \emph{overparameterized regime}, where the optimization is known to converge significantly faster, becoming less prone to local minima and alleviating the aforementioned issues \cite{kiani_learning_2020, kim_effectiveness_2021, larocca_control_2021, larocca_theory_2021}. Though significant progress in explaining the mechanism behind this phenomenon has been achieved \cite{xuchen_convergence_2022, larocca_theory_2021}, open questions remain, such as how to determine the optimal circuit depth that \emph{best} leverages this effect.

In this work, we begin by comprehensively characterizing the original QAOA protocol \cite{farhi_qaoa_2014} on a 1D lattice and a variation on it using decoupled angles \cite{herrman_multiangle_2021}. By deriving the \emph{explicit} structure of the associated Lie algebras \cite{larocca_theory_2021, morales_universality_2020}, we show that it can prepare \emph{exactly} all fermionic Gaussian states satisfying the symmetries of the circuit. Guided by these results, we find that, while decoupling the angles increases the number of parameters and the expressibility of the circuit by removing symmetries, this makes the preparation of non-local states \emph{easier} and of local states \emph{harder}, flipping the behavior observed when these angles are coupled. This contrasts with the commonly held belief that the use of symmetries is beneficial to the optimization \cite{barron_preserving_2021,gard_efficient_2020, 
larocca_group_2022,
meyer_symmetry_2022,
otten_symmetry_2019,
seki_symmetry_2020, shaydulin_symmetry_2021, zhang_shallow_2021, tilly_vqe_2021}.

By leveraging covariance matrix and automatic differentiation techniques to simulate deep circuits and system sizes up to $80$ qubits, we study the overparameterized regime of optimization, exploiting the fact that its onset is polynomial in lattice size for the ans\"atze we consider (see \cite{larocca_control_2021,larocca_theory_2021} and Section~\ref{sec:pstar}). In these conditions, we observe that the number of iterations to converge to the solution scales \emph{linearly} with the size of the system. This contrasts with the polynomial scaling we encounter at smaller depths. Moreover, we quantify the degree of improvement that overparameterization confers to the optimization as the circuit depth increases. We find that the number of iterations to converge decays \emph{exponentially} with circuit depth, and that the optimization hardness develops a minimum for a depth that is proportional to $L^2$, the square of the system size. Finally, we find that the exponential convergence to the solution seen in the overparameterized regime can be explained in terms of an improvement of the quality of the linear approximations provided by the gradients, elucidating why the optimization hardness continues to decrease even after local minima disappear.

This paper is organized as follows. 
In Sec.~\ref{sec:prelim} we introduce the variational setup, focusing on free-fermion systems, and establish the associated notation. 
Moreover, we develop Lie theoretical tools for the analysis of variational algorithms. We apply these tools in Sec.~\ref{sec:results} which contains our main results. In particular, Section~\ref{sec:pstar} characterizes the expressibility of protocols introduced in Section~\ref{sec:free_fermions}, where they are presented alongside a concise introduction to free fermionic systems. In Section~\ref{sec:traces}, we explore the optimization associated to the corresponding parameterized circuits, and highlight the influence of symmetries and the locality of the target state. In Section~\ref{sec:overparameterized} we discuss the effect of overparameterization, and perform a scaling of the number of iterations to converge with system size and circuit depth. Our conclusions are presented in Section~\ref{sec:conclusion}, while the Appendixes contain further numerical characterization of the circuit's optimization landscape in terms of gradient variance, staircase structure in optimization traces, and the details of the underlying Lie algebra structure.

\section{Preliminaries}\label{sec:prelim}
\subsection{Variational Quantum Algorithms} \label{sec:vqa}

Variational quantum algorithms (VQA) \cite{cerezo_variational_2021, benedetti_ml_2019} are generally formulated as a feedback loop between an optimization routine running on a classical computer, and a quantum simulator. This routine manipulates a set of controllable parameters defining a family quantum circuits with the objective of finding a circuit which is able to prepare a quantum state of interest (though more general definitions exist depending on the problem at hand). In this context, the parameterized family of quantum circuits we will focus on is constructed by following the alternating ``bang-bang" structure of the Quantum Approximate Optimization Algorithm (QAOA) \cite{farhi_qaoa_2014, yang_pontryagin_2017}. Given a set of Hamiltonians $\mathcal{H} = (H_1,...,H_m)$, the circuit is defined by the unitary operator
\begin{align}
	\nonumber U(\bm{\theta}, p) = \exp(- i \theta_{p,m} H_m) ... \exp(- i \theta_{p,1} H_1)  \\ ...\exp(- i  \theta_{1,m} H_m) ... \exp(- i \theta_{1,1} H_1). \label{eq:general_ansatz}
\end{align}
where $p$ controls the circuit depth and $\bm{\theta}\equiv \{ \theta_{1,1},\theta_{1,2},\ldots,\theta_{p,m} \}$ are the parameters to be optimized. We call such a set of Hamiltonians $\mathcal{H}$ a \emph{protocol}. The associated parameters are often called \emph{angles} in the literature.

Given an initial state $\ket{\psi(0)}$ and a set of parameters $\bm{\theta}$, this circuit prepares the state 
\begin{align} \label{eq:preparation}
	\ket{\psi(\bm{\theta})} = U( \bm{\theta}, p) \ket{\psi(0)}.
\end{align}
The goal, as outlined above, is to employ a classical optimization routine in order to find a set of angles $\bm{\theta}^*$, such that $\ket{\psi(\bm{\theta}^*)}$ is the quantum state of interest, which we call the \emph{target state}. This is done by supplying the optimizer with a cost function, which measures the distance between the state in preparation and the target state.

Here, we define the target state as being the unique ground state of some quantum Hamiltonian $H$. In this case, one can use a cost function based on the expectation value of the state with respect to this Hamiltonian. We use the shifted energy density as the cost function, 
\begin{align} \label{eq:energy_density}
  e(\psi) = \frac{\langle \psi|H| \psi \rangle - E_0}{L},
\end{align}
where $E_0$ is the ground state energy of $H$ and $L$ is the size of the system under consideration.  By the variational principle, and given that we assume that the target state is the unique ground state of $H$, this cost is minimized and is equal to zero when $\ket{\psi}$ is the target state. In what follows, we refer to the Hamiltonian featuring in this cost function as the \emph{target Hamiltonian}.

\subsection{Lie Theory and Expressibility} \label{sec:lie_theory}

\begin{figure}[hbt]
	\centering
	\includegraphics[width=\linewidth]{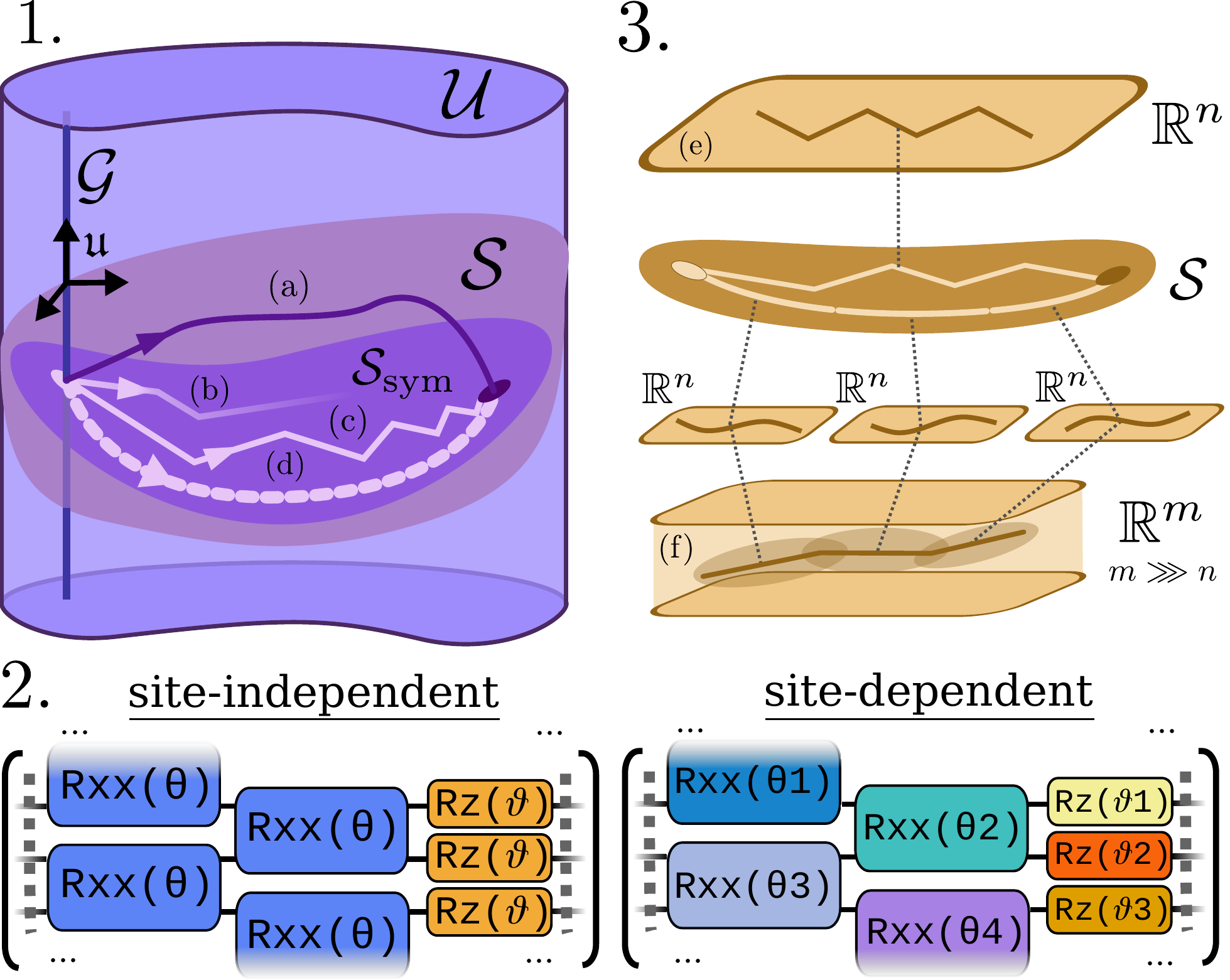}
	\caption{1. Schematic depicting the Lie structures introduced in Section~\ref{sec:lie_theory} and their relation to variational optimization. From an initial state, the set of unitaries generated by the parameterized circuit, $\mathcal{U}$, prepares a manifold of states $\mathcal{S}$. The space of directions that the protocol is able to explore at a given point is characterized by the Lie algebra $\mathfrak{u}$, and there is a redundancy in the unitaries preparing a state which is represented by a stabilizer gauge group $\mathcal{G}$. Symmetries in the protocol constrain the optimization (a) to a submanifold of states $\mathcal{S}_\text{sym}$, which affects the landscape by introducing local minima (b) and restricting the features available to the optimizer, causing the optimization to take longer (c), as explained in Section~\ref{sec:traces}. By increasing the depth of the circuit, the variational protocol enters an overparameterized regime (d), where minima disappear and convergence of the cost function to a global minimum becomes exponential. 2. Circuits illustrating the variational protocols introduced in Section~\ref{sec:free_fermions} and studied throughout the rest of this paper 3. The parameterization at minimal circuit depth $\hat p$ (e) matches the dimension of the manifold of states, so that minima exist and are unique. By the implicit function theorem, the overparameterization (f) defines local, lower-dimensional parameterizations matching the dimension of the manifold. These more adequately capture its features, while describing the optimization path in a piecewise manner (see Section~\ref{sec:overparameterized}). It is an open question whether a \emph{global} exact parameterization with these properties could be found. }
	\label{fig:schematic}
\end{figure}

Lie theoretical techniques provide a powerful tool to characterize the expressibility of quantum controllable systems \cite{albertini_controllability_2001}. They have been previously used in the literature to prove the universality of a set of protocols under certain assumptions \cite{morales_universality_2020}. These have been further employed to characterize controllability and overparameterisation in variational quantum algorithms \cite{larocca_control_2021,larocca_theory_2021}, and have recently found use in characterizing symmetries in data in the context of machine learning \cite{meyer_symmetry_2022, larocca_group_2022}. 

Here we summarize and formalize the theory behind this and define the associated notation. Given $U(\bm{\theta}, p)$ defined by the Hamiltonians $\mathcal{H} = (H_1,...,H_m)$ as in Section~\ref{sec:vqa}, we define
\begin{align}
&\mathcal{U}= \bigcup_{p=1}^\infty \mathcal{U}^p, 
  &\mathcal{U}^p = \{U(\theta, p) : \theta \in \mathbb{R}^{mp} \}.
\end{align}
The set $\mathcal{U}$ contains all unitaries that the protocol $\mathcal{H}$ can generate at arbitrary circuit depth. It is a group, as it contains the product of any two of its elements and the inverse of any of its elements. Further, since matrix multiplication is differentiable, it constitutes a \emph{Lie group}. 

Associated to the Lie group $\mathcal{U}$ is a \emph{Lie algebra} $\mathfrak{u}$, which can be defined at a point $\theta$ as
\begin{align}
    \mathfrak{u} = \left \{\frac{\partial U( \theta, p)}{\partial \theta_i} : i, p \in \mathbb{N}^+ \right \}.
\end{align}
It characterizes how the circuit $U(\theta,p)$ changes with an infinitesimal variation of the parameters. Note that $\mathcal{U} = \{e^{-iH} : iH \in \mathfrak{u} \}$ and that $\mathfrak{u} = \langle iH_1,...,iH_m \rangle$ \cite{morales_universality_2020, domenico_control_2007}, where $\langle ... \rangle$ denotes the Lie algebra generated by these elements i.e the space obtained by iteratively taking the Lie bracket $[A,B] = AB - BA$ of $iH_1,...,iH_N$ \cite{hall_lie_2003}. 

We further define
\begin{align}
&\mathcal{S}^p = \{U(\theta,p)|\psi(0)\rangle : \theta \in \mathbb{R}^{mp}\}, 
   &\mathcal{S} = \bigcup_{p=1}^\infty \mathcal{S}^p, 
\label{eq:states_manifold}  
\end{align}
as the set of states preparable by the variational quantum circuit at depth $p$ and at any depth, respectively. We emphasize that  $\mathcal{S}$ depends on the initial state chosen. 

We will restrict our analysis to finite dimensional spaces, for which there must exist a $p^*$ such that $\mathcal{U} = \mathcal{U}^{p^*}$ \cite{domenico_control_2007}.  By the same argument, there must be a $\hat p$ such that $\mathcal{S} = \mathcal{S}^{\hat p}$. This represents the circuit depth at which the circuit has reached \emph{maximum expressibility} for a given initial state. We will see that, in general, $\hat p < p^*$. This happens because there can be a set $\mathcal{G}$ of unitary matrices in $\mathcal{U}$ that leave the initial state $\ket{\psi(0)}$ invariant, forming a stabilizer subgroup. Mathematically, this  translates into $\mathcal{U}$ having a fiber bundle structure and $\mathcal{S} \cong \mathcal{U} / \mathcal{G}$, where $\mathcal{G}$ represents a \emph{Gauge symmetry group} \cite{nakahara_geometry_2017} (see Figure~\ref{fig:schematic}.1 for a graphical summary). As a consequence of this, 
\begin{align} \label{eq:s_dim}
  \dim \mathcal{S} = \dim \mathcal{U} - \dim \mathcal{G}
\end{align}
justifying that, in general, $\mathcal{S}$ will require less parameters to describe than $\mathcal{U}$. 
It is an open question whether there is a method to systematically determine $p^*$ and $\hat p$ given a set of Hamiltonians $\mathcal{H}$.

Since $\mathcal{U} = \{e^{-iA_1}e^{-iA_2}....e^{-iA_m} : m \in \mathbb{N}, A_j \in \mathfrak{u}\}$ \cite{morales_universality_2020, domenico_control_2007}, the unitaries in $\mathcal{U}$ can approximate a quantum annealing protocol and thus adiabatically prepare the ground state of any Hamiltonian in $\mathfrak{u}$ \cite{mbeng_annealing_2019}, provided that $\ket{\psi(0)}$ is the ground state of some $H_0 \in \mathfrak{u}$. Conversely, if $\ket{\psi}$ is prepared by $U \in \mathcal{U}$, then it is the ground state of $H = U H_0 U^\dagger \in \mathfrak{u}$. Thus, the Lie algebra $\mathfrak{u}$ fully characterizes the set $\mathcal{S}$ of preparable states; we will exploit this in Section~\ref{sec:pstar} to study the expressibility of the protocols defined in Section~\ref{sec:free_fermions}. 

\subsection{Free fermionic systems and VQA} \label{sec:free_fermions}

Throughout this paper, we work with a $1$D spin system on a linear lattice, and we denote the corresponding lattice size by $L$. We define Majorana operators through the standard Jordan-Wigner transformation
\begin{align}
  &\gamma_{2j-1} = Z...ZX_j,
  &\gamma_{2j} = Z...ZY_j,
\end{align}
where we assume that the string of $Z$s stretches from the left end of the lattice to the $i$th position.  Quadratic fermionic Hamiltonians (which we abbreviate to ``quadratic Hamiltonians") are of the form
\begin{equation}
  H = i \sum_{j, k} h_{j,k} \gamma_j \gamma_k,  
\end{equation}
where $h_{j,k}$ is a $2L\times 2L$ real and antisymmetric matrix. Note that all quadratic Hamiltonians preserve (commute with) the fermionic parity $P = \prod_j Z_j$. The eigenstates of a quadratic Hamiltonian are \emph{fermionic Gaussian states} (FGS), which are a class of quantum states that are fully determined (up to a phase) by their $2L\times 2L$ covariance matrix
\begin{align}
  \Gamma_{jk} &= \frac{i}{2} \langle \gamma_j \gamma_k - \gamma_k \gamma_j \rangle_\psi \nonumber \\ 
              &= \frac{i}{2} \trace (| \psi \rangle \langle \psi |(\gamma_j \gamma_k - \gamma_k \gamma_j)),
\end{align}
which is real and antisymmetric.

Both FGS and quadratic Hamiltonians are efficiently representable, requiring a number of parameters that is quadratic in system size to be specified. Moreover, the quantum dynamics associated to the evolution of a covariance matrix under the action of a quadratic Hamiltonian is efficiently computable, as is its expectation value with respect to a quantum observable \cite{valiant_quantum_2002, terhal_classical_2002, surace_fermionic_2021}.

We now establish how symmetries in the 1D lattice are reflected on the structures we have just introduced. The covariance matrix of a translationally invariant FGS $\ket{\psi}$ and a translationally invariant quadratic Hamiltonian defined by $h_{j,k}$ satisfy, respectively:
\begin{align}
  &\Gamma_{jk} = \Gamma_{j+2m \; k+2m}, 
  &h_{j,k} = h_{j + 2m, k + 2m}, \label{eq:ti_constraint}  
\end{align}
for all integers $m$ and it is understood that coefficients are taken modulo the lattice size. The covariance matrix of a lattice inversion symmetric FGS $\psi$ and a lattice inversion symmetric quadratic Hamiltonian $H$ defined by $h_{j,k}$ satisfy, respectively:
\begin{align}
  &\Gamma_{jk} = (-1)^{j-k+1}\Gamma_{L-k+1 \; L-j+1}, \nonumber \\
  &h_{j,k}  = (-1)^{j-k+1} h_{L-k+1 \; L-j+1} \label{eq:li_constraint}
\end{align}
Throughout, we denote by ``symmetric FGS" and ``symmetric quadratic Hamiltonians"   those that are invariant both under translation and lattice inversion. 

Following the framework outlined in Section~\ref{sec:vqa} we study two different protocols, both of which feature only quadratic Hamiltonians:
\begin{enumerate}
\item A \emph{site-independent} protocol, defined by the set
\begin{align} \label{eq:site_independent_protocol}
\nonumber \mathcal{I} &= \left(i \sum_j Z_j, i \sum_j X_j X_{j+1}\right)  \\ 
&= \left(-\sum_j \gamma_{2j-1} \gamma_{2j}, -\sum_j  \gamma_{2j} \gamma_{2j+1}\right). 
\end{align} 
and we denote its Lie algebra by $\mathfrak{i}$. 
\item A \emph{site dependent} protocol, where 
\begin{align} \label{eq:site_dependent_protocol}
  \mathcal{D} = (&iX_1 X_2, ..., iX_{N-1}X_N, \nonumber \\
              &iZ_1, ..., iZ_N)               \nonumber \\
              = (&-\gamma_2 \gamma_3, ..., -\gamma_{N-1} \gamma_{N-2}, \nonumber\\
              &-\gamma_1 \gamma_2, ..., -\gamma_{N-1} \gamma_{N}),
\end{align}
and we denote its Lie algebra by $\mathfrak{d}$. 
\end{enumerate}
The former corresponds to the original QAOA protocol \cite{farhi_qaoa_2014} on a linear lattice, while the latter results from removing the layer-wise coupling in the angles of this original protocol \cite{herrman_multiangle_2021}. We will see that this decoupling results in distinct properties with respect to the optimization of the associated variational algorithm.

Writing out the corresponding unitary explicitly:
\begin{widetext}
\begin{align}
	U(p, \theta) = \exp(- i \sum_k \theta_{p,Z}^k Z_k) \exp(- i \sum_k  \theta_{p,XX}^k X_kX_{k+1}) ...
  \exp(- i \sum_k \theta_{1,Z}^k Z_k)  \exp(- i \sum_k \theta_{1,XX}^k X_kX_{k+1}). \label{eq:ansatz}
\end{align}
\end{widetext}
As mentioned above, the site-independent protocol can be seen as the site-dependent one with the additional constraint that $\theta^i_{a,P} = \theta_{a,P}$, i.e., the value of the angle is the same across a circuit layer. 

The full structure of the Lie algebras corresponding to the protocols above is elucidated in Appendix~\ref{sec:lie}. In particular, a basis for these algebras is obtained by iteratively taking the commutators of their generators.  Moreover, we clarify how their structure changes when we restrict ourselves to a particular sector of the parity symmetry.
In the next section, we will employ them to study the expressibility of the corresponding protocols following Section~\ref{sec:lie_theory}.

\section{Results}\label{sec:results}

\subsection{Circuit expressibility and saturation} \label{sec:pstar}

Here, we determine the expressibility of the protocols introduced in the previous section by examining the corresponding Lie algebra. In particular, we study the set of unitaries $\mathcal{U}$ that each protocol can generate and the set of states $S$ that each can prepare. In what follows we consider the initial state to be a fermionic Gaussian state of a given parity respecting the symmetries of the circuit.

Our results are summarized in Table~\ref{table:expressibility}. All protocols are able to prepare every FGS with the same parity as the initial state and respecting the symmetries of the circuit. This can be proved by studying the structure of the corresponding Lie algebras, derived in Appendix~\ref{sec:lie} and referenced in Table~\ref{table:expressibility}. By applying the Jordan-Wigner transformation, these Lie algebras can be seen to form the set of free fermionic Hamiltonians satisfying the symmetries of the circuit. As shown in Section~\ref{sec:lie_theory}, every ground state of such a Hamiltonian can be prepared by the circuit at some depth; and the set of these ground states are precisely the fermionic Gaussian states having the same parity as the initial state and respecting the appropriate symmetries.

As outlined in Section~\ref{sec:lie_theory}, there can be a symmetry subgroup $\mathcal{G}$ of $\mathcal{U}$ that leaves the initial state invariant. In the case of a FGS, there is a $U(L)$ freedom in the fermionic modes, which can be rotated without changing the underlying state \cite{windt_local_2021}. The subset of these rotations contained in $\mathcal{U}$ will form $\mathcal{G}$; this can be all of the $U(L)$ freedom, as is the case of the site dependent protocol, or only part of it, in which case $\dim \mathcal{G} < \dim U(L)$.

We further attempt to determine the minimum depth $\hat p$ necessary to prepare any state in $\mathcal{S}$ for each of the cases in Table~\ref{table:expressibility}. In what follows, we denote by $q$ the number of variational parameters per unit of $p$. While it is clear that $q \hat p$ must be greater than $\dim \mathcal{S}$, the circuit must be also be deep enough so that correlations are able to propagate across the lattice \cite{mbeng_annealing_2019}. A consequence of this is that $\hat p \geq \lceil L/2 \rceil$. We compute $\hat p$ numerically by randomly generating Hamiltonians in $\mathfrak{u}$ and verifying that their ground states are prepared to numerical precision. The quantity
\begin{align}
q\hat p - \dim \mathcal{S},
\end{align}
represents the number of parameters in the circuit that exceeds $ \dim \mathcal{S}$. 

We find, in cases where periodic boundary conditions (PBC) are employed, or where the site-dependent protocol is used, that $\hat p = \lceil L / 2 \rceil$, saturating the aforementioned lower bound. For the remaining case, which corresponds to the site-independent protocol using open boundary conditions (OBC), $\hat p$ proved to be unfeasible to determine numerically in a precise manner due to a significantly higher number of local minima.

In the site-independent case with periodic boundary conditions, $q \hat p = \dim \mathcal{S} $, which suggests an \emph{exact parameterization} of $\mathcal{S}$. In contrast, in the site-dependent case with PBCs, there are $q \hat p - \dim \mathcal{S} = L$ redundant parameters at $\hat p$. One can, however, do away with them by removing the last $e^{-i\sum_j \theta_{p,Z} Z_j}$ layer from this circuit, while still being able to prepare all states in $\mathcal{S}$. Thus, in this case, one can also obtain an exact parameterization. 
By abuse of terminology, we will refer to the behavior at circuit depth $\hat p$ as the \emph{exactly parameterized regime}, regardless of whether $q\hat p = \dim \mathcal{S}$. A consequence of having an exact parameterization of $\mathcal{S}$ is that, when the associated angles are appropriately restricted, global minima exist and are unique.

\begin{table}[htb]
  \centering
  \setlength\tabcolsep{1.5pt} 
  \resizebox{\columnwidth}{!}{%
  \begin{tabular}{ |c|c|c|c|c|  }
   \hline
   & \multicolumn{2}{|c|}{Dependent} & \multicolumn{2}{|c|}{Independent} \\
   \hline
   & OBC & PBC & OBC & PBC\\
   \hline
   \hline
   \hline
   $\mathcal{S}$ & \multicolumn{2}{|c|}{fixed parity FGS} & \thead{fixed \\ parity \\ FGS \\ satisfying \\ \eqref{eq:li_constraint}} & \thead{fixed \\ parity \\ FGS \\ satisfying \\ \eqref{eq:ti_constraint} \& \eqref{eq:li_constraint}}\\
   \hline
   \hline
   $\mathfrak{u}$ & \eqref{eq:dep_algebra_obc} & \eqref{eq:dep_algebra_pbc} & \eqref{eq:indep_algebra_obc} & \eqref{eq:indep_algebra_pbc} \\
   \hline
   \thead{$\mathfrak{u}$ \\ (fixed parity)} & \multicolumn{2}{|c|}{\eqref{eq:dep_algebra_obc}} & \eqref{eq:indep_algebra_obc} & \eqref{eq:parity_indep_algebra_pbc} \\
   \hline
   \hline
   $\dim \mathcal{U}$ & $L(2L-1)$ & $2L(2L-1)$ &$L^2$& $3L - 2$ \\
   \hline
   \thead{$\dim \mathcal{U}$ \\ (fixed parity)} & \multicolumn{2}{|c|}{$L(2L-1)$} & $L^2$ & $\lfloor 3L/2 \rfloor $\\
   \hline
   $\dim \mathcal{G}$ & \multicolumn{2}{|c|}{$L^2$} & $L(L+1)/2$ & $\lfloor L/2 \rfloor $\\
   \hline
   $\dim \mathcal{S}$ & \multicolumn{2}{|c|}{$L(L-1)$} & $L(L-1)/2$ & $L$ \\
   \hline
   \hline
  \end{tabular}
  }
  \caption{Table summarizing the expressibility of the site-dependent, Eq.~\eqref{eq:site_independent_protocol}, and site-independent, Eq.~\eqref{eq:site_dependent_protocol}, protocols. The structure of the Lie algebras $\mathfrak{u}$ corresponding to each protocol is given in Appendix~\ref{sec:lie}; the entries in this table refer to the respective basis. As outlined in the main text, these are used to analytically deduce $\mathcal{U}$, the space of unitaries that each protocol can generate, and $S$, the space of states that each protocol can prepare.  We assume that the initial state is a FGS of a given parity respecting the symmetries of the circuit.
  }
  \label{table:expressibility}
\end{table}

\subsection{Effect of Symmetries and Locality on the Optimization Landscape} \label{sec:traces}

We proceed to study the hardness of the optimization and the characteristics of the associated landscape when running a variational algorithm using the site-independent, Eq.~\eqref{eq:site_independent_protocol}, and site-dependent, Eq.~\eqref{eq:site_dependent_protocol}, protocols. We work with PBC, and target the ground state of two models:
\begin{enumerate}
  \item The critical transverse field Ising model~\cite{ho_hsieh_2019}
  \begin{equation} \label{eq:ising}
    H_I = - \sum_{j} X_j X_{j+1} - \sum_j Z_j.
  \end{equation}
  \item Randomly generated symmetric quadratic Hamiltonians
  \begin{equation} \label{eq:random}
    H_\text{G} = i \sum_{jk} {h}_{jk} \gamma_j \gamma_k
  \end{equation}
  where ${h}_{jk}$ respects Eq.~\eqref{eq:li_constraint} and Eq.~\eqref{eq:ti_constraint}.
\end{enumerate}
The first is a well-known quantum-critical model in condensed matter physics~\cite{dutta_qpt_2015}, possessing a ground state whose entanglement entropy diverges logarithmically with system size \cite{Calabrese2004}. The second Hamiltonian is obtained by sampling at random out of all the ones for which the ground state is possible to prepare with both protocols. As mentioned at the end of Section~\ref{sec:free_fermions}, this is characterized by the Lie algebra corresponding to each protocol; in practice, the algebra of the site-dependent one contains that of the site-independent, and the latter, when using PBC, consists of all quadratic Hamiltonians satisfying Eqs.~\eqref{eq:ti_constraint}-\eqref{eq:li_constraint}. These Hamiltonians are sampled by directly generating entries in $h_{ij}$ using a normal distribution with mean equal to zero and standard deviation equal to one, following these constraints.

Moreover, below we use a $Z$-polarized state
\begin{align} \label{eq:initial_state}
	\ket{\psi(0)} = \ket{\uparrow ... \uparrow}
\end{align}
as the initial state of the protocol.
The classical minimization is performed using the BFGS optimization algorithm; though other optimizers such as Nelder-Mead and conjugate gradient were checked and the behavior obtained was qualitatively the same.

\begin{figure*}[hbt]
	\centering
	\includegraphics[width=\linewidth]{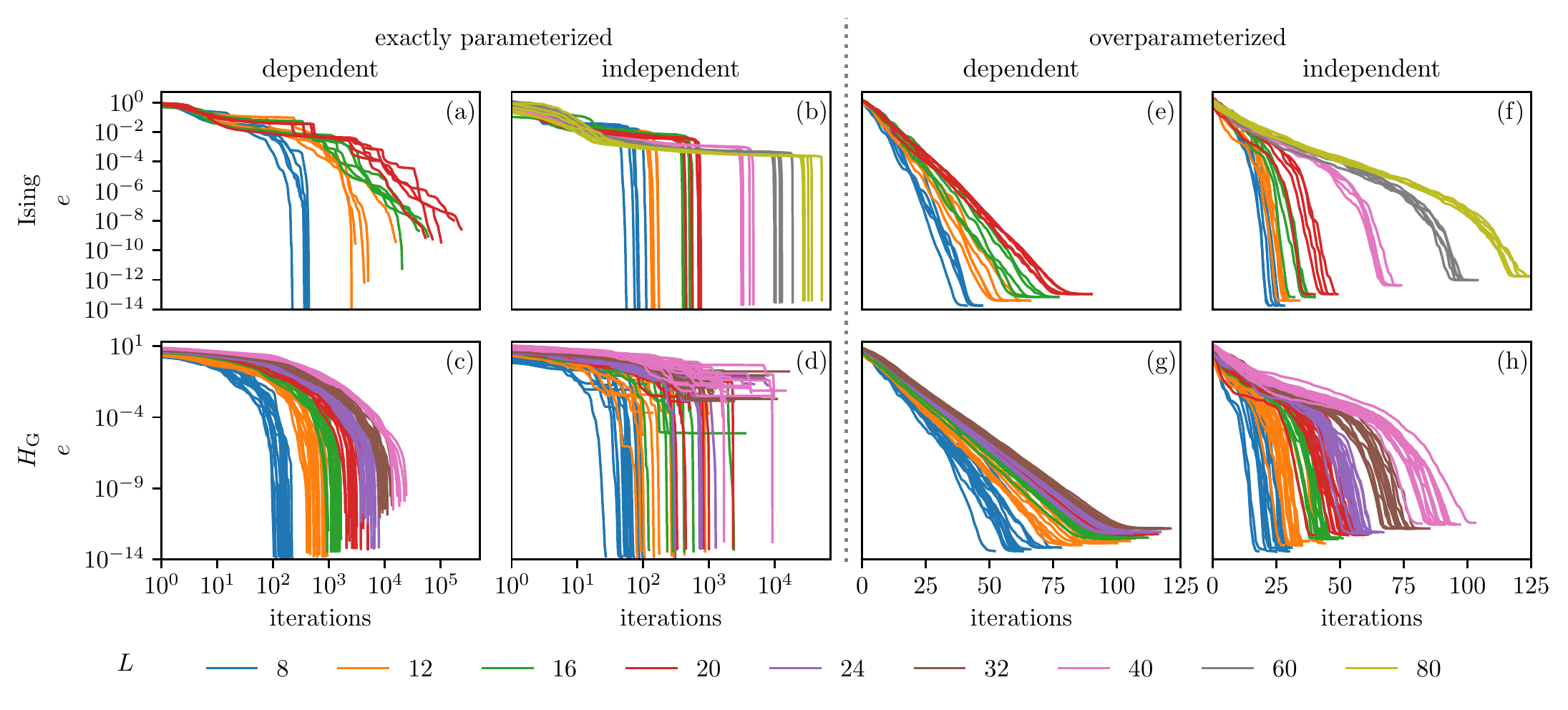}
	\caption{Cost function optimization traces exposing the differences between the minimizations as the protocol and target Hamiltonian change. The effect of circuit depth is probed using the exactly   parameterized regime ($p=L/2$) and the overparameterized regime ($p=L^2/4$). The target Hamiltonian is (Top row): the Ising model, defined in Eq.~\eqref{eq:ising}, as the target Hamiltonian, and $5$ random initializations per value of $p$, lattice size and protocol. (Bottom row): $3$ randomly generated symmetric quadratic Hamiltonians, defined in Eq.~\eqref{eq:random}, and $5$ random initializations per generated Hamiltonian, value of $p$, lattice size and protocol.
	When preparing the ground state of the Ising model (a), the site dependent protocol displays a "staircase" pattern, where the cost function undergoes little variation before dropping to a new plateau; in stark contrast, when preparing a generic symmetric FGS (c), it exhibits a smoother decrease. The site independent protocol presents the opposite behavior: when preparing the ground state of the Ising model (b), the cost initially undergoes a slow, but smooth, decrease, before sharply dropping when the state is prepared; when the target state is a generic symmetric FGS (d), the staircase pattern is again visible, this time also accompanied by local minima. This behavior is highlighted in Figure~\ref{fig:staircase} in Appendix~\ref{sec:overlaps}. After increasing the depth of the circuit into the overparameterized regime, the differences in optimization between states vanish, and the cost function decrease becomes exponential with no local minima present.
	}
	\label{fig:iter_vs_cost}
\end{figure*}

Figure~\ref{fig:iter_vs_cost} shows the optimization traces after classically optimizing the algorithm. We compare two circuit depths: $p=L/2$ \cite{ho_hsieh_2019, mbeng_annealing_2019, wang_maxcut_2018}, which we have determined to be the minimum depth for which the protocol reaches maximum expressibility, and $p=L^2/4$, well into the overparameterized regime (as we quantify in Section~\ref{sec:overparameterized}), where the redundancy in parameters is known to greatly reduce the computational cost of the optimization \cite{kiani_learning_2020, kim_effectiveness_2021, larocca_control_2021, larocca_theory_2021}. We defer a discussion of the latter for Section~\ref{sec:overparameterized}.

We observe that changing the target state and the protocol employed can drastically alter the characteristics of the optimization. In particular, we identify a ``staircase" pattern, signaling a harder optimization problem, which is discernible at higher system sizes. It emerges both when employing the site dependent protocol to target the Ising model [Figure~\ref{fig:iter_vs_cost}(a)], and when using the site independent protocol to target generic quadratic Hamiltonians [Figure~\ref{fig:iter_vs_cost}(d)].  We discuss and propose a mechanism for this phenomenon in Appendix~\ref{sec:overlaps}, where we see that the cost function is not able to distinguish the target state from other states in the Hilbert space, even if they are orthogonal. It was argued in  \cite{arrasmith_gorges_2021} that this behavior is not possible when the barren plateau phenomenon is present; here, we see that it can represent an intermediate behavior of the landscape as the size of the system increases and gradients begin to vanish (we study the scaling of the gradients in Appendix~\ref{sec:gradient}). The efficient classical simulation of FGS is pivotal in observing it, as it is not as evident in smaller system sizes.

The only case susceptible to local minima is the site independent one at $p=L/2$ when targeting generic symmetric quadratic Hamiltonians [Figure~\ref{fig:iter_vs_cost}(d)]. There, the optimization is highly sensitive to the initial condition, and the number of iterations to converge, along with the value found at the minimum, can vary drastically. This starkly contrasts with the other cases in Figure~\ref{fig:iter_vs_cost} --- in particular, the one where we target the Ising model using the same protocol [Figure~\ref{fig:iter_vs_cost}(b)]. While the average number of iterations to convergence is approximately the same in both cases, the former has a high standard deviation, as can be seen in Figure~\ref{fig:ls_vs_iter}. Thus, using a protocol with less symmetry produces more consistent results which are not susceptible to local minima. This shows that imposing more symmetries may not always be desirable, as it may have an unpredictable effect in the optimization landscape and can result in local minima.

We now study how different properties of the target Hamiltonian can give rise to the phenomenology we previously observed and thereby explain Figure~\ref{fig:iter_vs_cost}. One obvious property that distinguishes the Ising Hamiltonian, Eq.~\eqref{eq:ising}, from that of the random Hamiltonian in Eq.~\eqref{eq:random} is the presence of \emph{locality} in the former case. Locality of the target Hamiltonian is known to influence whether the optimization associated to a quantum circuit will feature barren plateaus, with non-local terms presenting exponentially vanishing gradients \cite{cerezo_cost_2021}.  Locality can also have an influence below system sizes at which barren plateaus appear, and it has been argued that long-range interactions in the target Hamiltonian make the optimization harder \cite{tang_longrange_2022}, resulting in higher values of the cost function at the optimum and requiring more iterations to converge. In our case,
we will see that the influence of the locality of the target Hamiltonian on optimization depends on the constraints of the protocol being used. In particular, we will show that the site-independent and the site-dependent protocol behave differently in this respect, which explains the observed differences in the optimization. 

We use three families of models to quantify how the locality of the target Hamiltonian affects the hardness of the optimization:
\begin{enumerate}
\item A special type of a long-range Ising Hamiltonian:
\begin{align}
  H(\alpha) &= - \sum_{r} e^{-\alpha r} \sum_j X_j Z_{j+1}Z_{j+2}\ldots Z_{j+r}X_{j+r+1} \nonumber \\ 
  &- \sum_j Z_j, \label{eq:longrange}
\end{align}
where $\alpha$ describes exponentially decaying interactions in a lattice. The choice of this Hamiltonian is motivated by the fact that its ground state can be expressed in terms of free fermions for any $\alpha$, unlike the related models with power-law decaying interactions recently studied in Refs.~\cite{ho_ultrafast_2019, tang_longrange_2022}.

\item $(k+2)$-local, symmetric, quadratic Hamiltonians
\begin{align} \label{eq:local_generic}
  H_\text{LG}(k) &= i \sum_{jl} \tilde{h}_{jl} \gamma_j \gamma_l, \\
  \tilde{h}_{jl} &= \begin{cases}
  \mathrm{random}, \; \mathrm{if} \; |j-l| < 2(k+2), \\
  0, \; \mathrm{if} \; |j-l| \geq 2(k+2),
  \end{cases}
   \nonumber
\end{align}
which are derived from the randomly-generated generic symmetric quadratic Hamiltonians in Eq.~\eqref{eq:random} by  setting $h_{jl}=0$ for any pair of Majoranas at a distance $\geq 2(k+2)$. 

\item A cluster Ising model at criticality \cite{doherty_identifying_2009, okada_identification_2022}
\begin{align} \label{eq:cluster}
  \nonumber H_\text{C}(k) &= - \sum_j X_j Z_{j+1}Z_{j+2}\ldots Z_{j+r}X_{j+k+1} \\
  &- \sum_j Z_j,
\end{align}
for which the ground state in one of the gapped phases is a symmetry-protected topological state \cite{raussendorf_cluster_2003}. 
\end{enumerate}
In the two latter models, interactions are strictly limited to sites at most $k+2$ sites away, while in the first model they are exponentially suppressed.

\begin{figure}[htb]
	\centering
	\includegraphics[width=\linewidth]{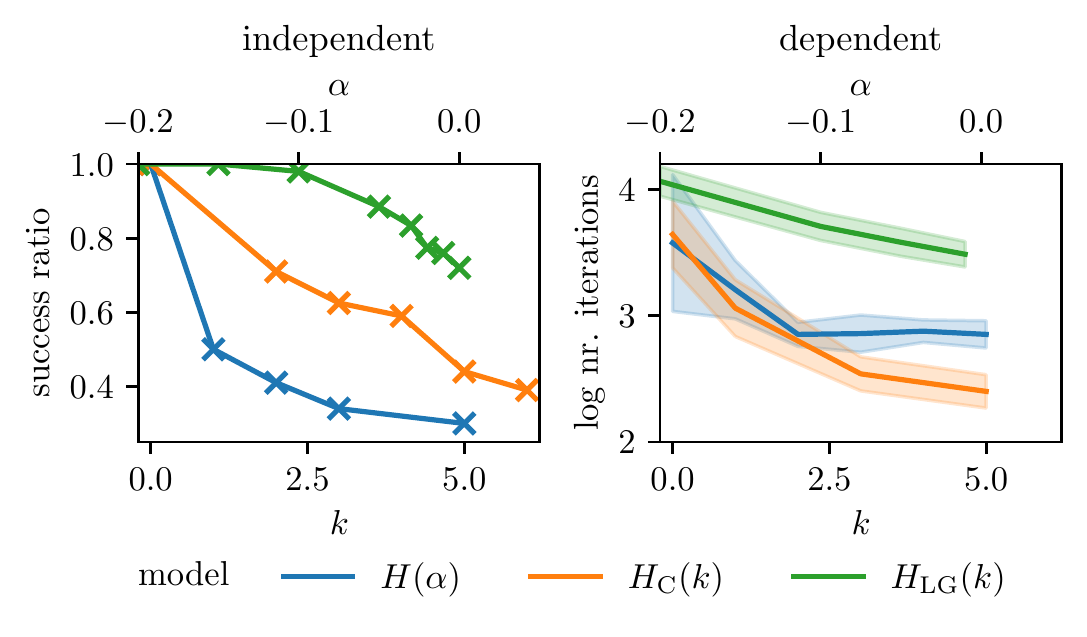}
	\caption{Plots characterizing the effect of Hamiltonian locality on optimization hardness in both the site-dependent and site-independent protocols in the exactly parameterized regime using PBCs. Horizontal axes are the parameters $k$ and $\alpha$, which control the locality of the Hamiltonians studied. On the left, the $y$- axis shows the success ratio, defined as the ratio between the number of random initializations which ended in state preparation and all random initializations. In the right panel, the $y$-axis is the (logarithm of) total number of iterations to converge. The labels refer to the target Hamiltonians defined in Eqs.~\eqref{eq:longrange}, \eqref{eq:local_generic} and \eqref{eq:cluster}. Lattice sizes used were either $12$ or $16$. Between 20 and 150 random initialisations were computed for each Hamiltonian parameter in the site dependent cases, and between 200 and 500 were computed in the site independent ones; solid line is the mean value, while the shaded area indicates standard deviation.
	}
  \label{fig:locality}
\end{figure}

Figure~\ref{fig:locality} compares the effect of locality on the optimization. We vary the parameters controlling localization of the couplings in each of the models Eqs.~\eqref{eq:longrange}-\eqref{eq:cluster}, and we measure the success probability in the site-independent case or the number of iterations to converge in the site-dependent case.  The success probability is defined as the ratio between the number of random initializations that resulted in the cost function dropping below numerical precision (and thus the target state being successfully prepared) versus the total number of initializations. This measure was not used as a benchmark for the site-dependent protocol, as this protocol is not susceptible to getting trapped in local minima, thus the success probability is always equal to one regardless of the locality of the Hamiltonian.

We see in Figure~\ref{fig:locality} that the more non-local the target Hamiltonian is, the lower the success probability is in the site-independent protocol. Surprisingly, however, we see that the more non-local the target Hamiltonian is, the \emph{lower} the number of iterations is to converge is the site-dependent case. Both statements are verified for all the models introduced above. Thus, while locality makes it \emph{easier} to prepare the target state using site-independent protocol, it makes the site-dependent protocol \emph{harder} to optimize.  We conclude that, on the one hand, the symmetry constraints in the site-independent protocol cause non-locality in the cost function to drive the optimization into difficult regions that trap it in local minima. The site-dependent case, on the other hand, is free to explore the entire manifold of FGS and bypass these traps, and non-local terms lead the optimization to converge faster, consistent with Figure~\ref{fig:schematic}(a), (b), (c).

\subsection{Overparameterized regime} \label{sec:overparameterized}

It has been pointed out~\cite{wiersema_hva_2020, kiani_learning_2020}  that taking the circuit depth to be very large, the optimization associated with Eq.~\eqref{eq:general_ansatz} becomes considerably easier -- a phenomenon dubbed  \emph{overparameterizaton}. 
The onset of the overparameterized regime has been argued to correspond the circuit depth at which the Quantum Fisher Information Metric saturates at every point $\theta$ in the optimization landscape \cite{larocca_theory_2021, capacity_haug_2021}. This is equivalent to the circuit depth at which an increase in $p$ does not lead to an increase in the states that can be prepared by the variational circuit Eq.~\eqref{eq:preparation}, i.e., the circuit depth corresponding to $\hat p$ as defined in Section~\ref{sec:lie_theory}. We numerically confirm this to be the case for all the cases discussed at the end of Section~\ref{sec:pstar}.

\begin{figure}[hbt]
	\centering
	\includegraphics[width=\linewidth]{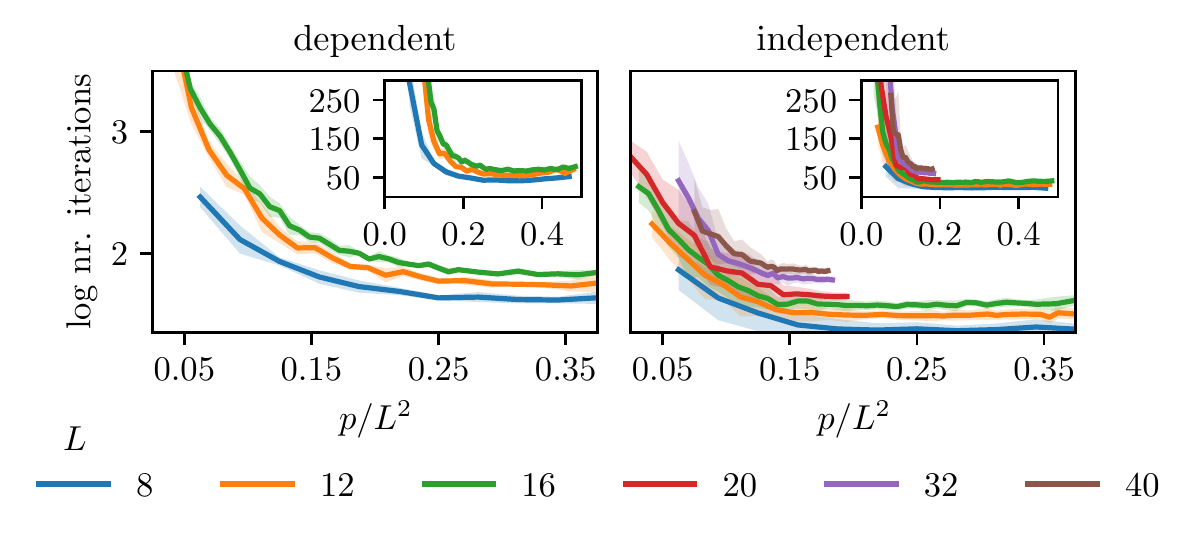}
	\caption{Number of iterations to converge vs. the circuit depth in the overparameterized regime. We see a decrease, consistent between different system sizes, when employing a circuit depth proportional to $L^2$. This reduction in the number of iterations is at first pronounced but then tapers off until it finally saturates. The insets use a linear scale for the vertical axis; they emphasize the point at which this saturation is reached. The target Hamiltonian is that of the Ising model, Eq.~\eqref{eq:ising}. Between 5 and 40 random initilisations were performed for each circuit depth; solid lines represent the mean value, and the shaded area indicates standard deviation.}
	\label{fig:overparameterized_scaling}. 
\end{figure}

\begin{figure}[hbt]
	\centering
	\includegraphics[width=\linewidth]{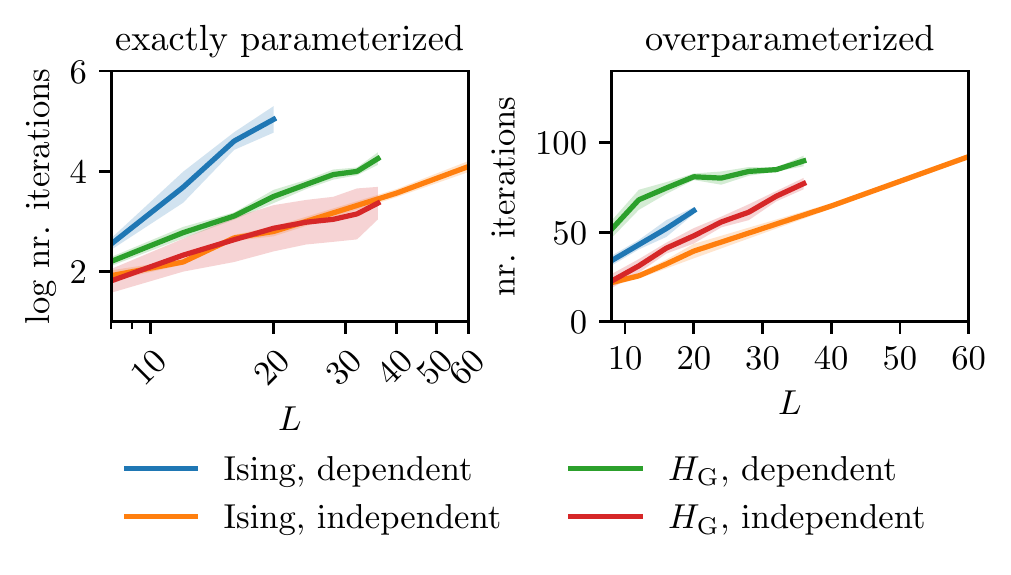}
	\caption{Number of iterations to converge vs. lattice size corresponding to the data in Figure~\ref{fig:iter_vs_cost}. Solid line is the mean value, the shaded area indicates the standard deviation. The legend refers to the Hamiltonian targetted and protocol used, in this order. The target Hamiltonians are either the Ising model,  Eq.~\eqref{eq:ising}, or generic symmetric quadratic Hamiltonians, Eq.~\eqref{eq:random}.	The number of iterations to converge scales polynomially with system size when $p=L/2$ (left), which is the minimum circuit depth for maximum expressibility. This turns into a linear scaling as the circuit enters the overparameterized regime (right).}
	\label{fig:ls_vs_iter}
\end{figure}

We perform a scaling analysis of the number of iterations that the optimizer takes to prepare the state as the depth of the circuit increases well into the overparameterized regime. We find that, as depth increases, the average number of iterations to converge initially suffers a large initial decay, until it slows down and saturates at $p\sim L^2$, i.e., no further increase in circuit depth provides a decrease in the average number of iterations to converge. We observe this trend consistently between different optimizers and different system sizes, the latter shown in Figure~\ref{fig:overparameterized_scaling}, where an exponential decay is seen. Further, by comparing how the average number of iterations the optimizer takes to converge scales with lattice size, both when the circuit depth is equal to $\hat p$ and into the overparameterized regime, we see that what is initially a polynomial scaling turns into a linear scaling with lattice size -- see Figure~\ref{fig:ls_vs_iter}.

From the above, we note that the overparameterized regime effectively represents a shift in the workload from the quantum computer to the classical one and vice-versa, as increasing the number of parameters makes the classical optimization easier, but the preparation of the state in the quantum computer harder given the increased circuit depth (and corresponding time to run the circuit, see Appendix~\ref{sec:hessian}). Thus, as noise levels in a device decrease, increasing the depth of the circuit allows variational algorithms to take immediate advantage of these advancements. This has the caveat that when the number of parameters passes a certain threshold, the overhead associated with certain algorithms, such as BFGS, exceeds the advantage obtained from overparameterizing the circuit; we quantify this in Appendix~\ref{sec:hessian}. In this case, algorithms designed to handle a large number of parameters, such as ADAM or stochastic gradient descent, should be employed instead.

Since in the site dependent case each circuit layer has $\sim L$ parameters, the behavior we have just described in terms of circuit depth $p \sim L^2$ corresponds to $\sim L^3$ parameters in this case. As all quantities in Table~\ref{table:expressibility} scale quadratically or linearly with system size, the phenomenon of overparameterisation can neither be explained by the saturation of the manifold of preparable states nor by the saturation of the manifold of unitaries. In what follows, we propose and test an explanation for gradient based optimizers in terms of a change in the very properties of the parameterisation of the manifold as the circuit depth increases.

A gradient based optimizer is an algorithm that, given an initial condition $\theta_0$, iterates the following update function
\begin{align*}
  \theta_{i+1} = \theta_i - \eta A \nabla e(\theta_i)
\end{align*}
until it converges, that is, it can not find a value of $\eta$, called the \emph{learning rate}, such that the update reduces the cost function $e$. The matrix $A$ is a bias that provides extra information to the algorithm. It can be the inverse of the Hessian $H^{-1}$ in the case of Newton based methods (or an approximation of it as in the case of quasi-Newton methods such as BFGS), or the inverse of the metric of the manifold being optimized over, as is is done in e.g. Quantum Natural Gradient descent methods \cite{stokes_qng_2020}. Importantly, the gradient of a function is a linear local approximation of the function at that point. While that means that, if $\norm{\nabla e} \neq 0$, there is a value of $\eta$ such that $e(\theta_{i+1}) < e(\theta)$, it does not offer any real guarantee about the actual change $\Delta e_{i+1} = e(\theta_{i}) - e(\theta_{i+1})$. 

\begin{figure}[tbh]
	\centering
	\includegraphics[width=\linewidth]{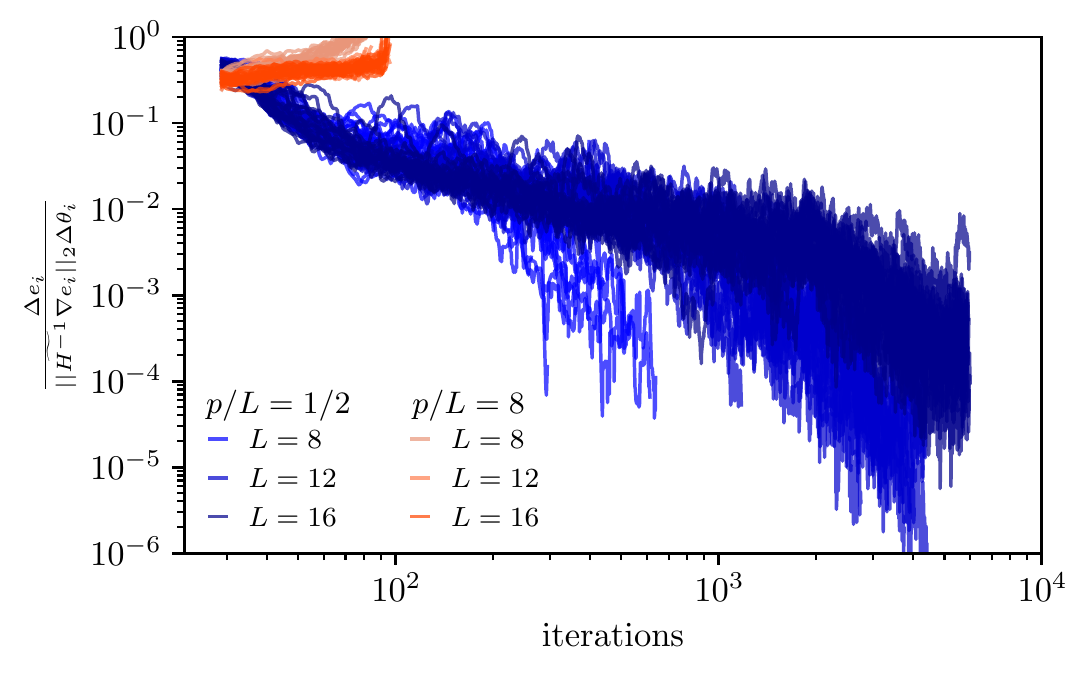}
	\caption{Quantification of how well the gradient accounts for the change in the cost function along the optimization, both in the exactly parameterized ($p=L/2$) and in the overparameterized ($p/L=8$) regimes. The plot shows the quantity defined in Eq.~\eqref{eq:gradient_quantifier} recorded throughout the optimization. We see that in the overparameterized regime, the value of the gradient consistently predicts the decrease in the cost function up to a constant factor; while it only accounts for a decreasing fraction of this variation in the exactly parameterized regime. Here, the target state was that of the Ising model with PBCs.}
	\label{fig:linear_param}
\end{figure}

Here, we examine how good this local approximation is as the optimization progresses, both when the circuit depth is equal to $\hat p$ and in the overparameterized regime. We run the BFGS algorithm, and the learning rate is picked on a per-iteration basis by using the strong Wolfe conditions, an established heuristic based on a minimum descent criterion~\cite{nocedal_numerical_2006}. This is a quasi-Newton method, and so $A = \widetilde{H^{-1}}$ will be an approximation to the inverse of the Hessian. In Figure~\ref{fig:linear_param}, we plot
\begin{align} \label{eq:gradient_quantifier}
  l = \frac{\Delta e_i}{||\widetilde{H^{-1}} \nabla e_i||_2 \Delta \theta_i},
\end{align}
which quantifies how much of the variation in the cost function can be attributed to the local approximation given by the gradient at the $i$th iteration. We see that in the overparameterized regime, most of the variation in the cost is accounted for by this approximation, and that it concentrates around an average of $\tilde k$ leading to the exponential decay seen in Figure~\ref{fig:iter_vs_cost} following the equation ${\partial e}/{\partial \theta} = \tilde k e$. We conclude that the overparameterized regime leads to parameterizations that are more amenable to optimization, as they capture the variation in the cost for longer distances in the parameter space (see panel 3 in Figure~\ref{fig:schematic}).

\section{Conclusion} \label{sec:conclusion}

By deriving the corresponding Lie algebra structure, we showed that the original QAOA protocol on a 1D lattice can prepare all fermionic Gaussian states satisfying the symmetries of the circuit, and we have numerically determined the circuit depth needed to achieve maximum expressibility. The efficient classical simulation of these states was employed to systematically study the optimization associated to these protocols.

We observed that decoupling the angles of the protocol makes the preparation of non-local states \emph{easier}, and of local states \emph{harder}, which is the opposite of what is observed when the angles are coupled. We argued that this is due to the symmetries in the system constraining the features available to the optimizer.
Further, we studied in detail the overparameterized regime, exploiting the larger system sizes and circuit depths accessible to us, where we find that the number of iterations to converge to the solution scales linearly with system size. Moreover, we found that the number of iterations to converge to the solution decreases exponentially with the depth of the circuit, until it saturates at a depth which is quadratic in system size. Finally, we observed that the improvement in the optimization can be explained in terms of of better local linear approximations provided by the gradients.

Beyond elucidating the current knowledge on the interplay between symmetry and optimization, and furthering the understing on the overparameterized regime of optimization, our work can serve as the basis for a benchmarking scheme for the implementation of variational algorithms in quantum computers. These have already been performed using free states such as, e.g., Majorana zero modes \cite{sung_preparing_2022} or the ground state of the Ising model \cite{alba_ising_2018}; this scheme could potentially be leveraged into error correcting methods \cite{sanisic_strategies_2020}. Moreover, it provides a framework to better understand the theory behind the preparation of free states using variational algorithms, already studied in  models such as the Ising model \cite{ho_hsieh_2019,  dreyer_critical_2021}, the Kitaev model in the exactly solvable limit \cite{jahim_kitaev_2022},
or the cluster model \cite{okada_identification_2022}. Furthermore, while there are established algorithms to build circuits that prepare FGS \cite{jiang_quantum_2018, kivlichan_simulation_2018}, these require a full description of the corresponding covariance matrix; a variational approach is relevant where this structure is not known beforehand e.g. when approximating interacting states \cite{okada_identification_2022, matos_quantifying_2021} or maximizing a quantity of interest such as magic \cite{hebenstreit_magic_2019, fermion_sampling_2022}. 
Finally, by fully characterizing these circuits on 1D lattices, we open the possibility to describe more complex graphs in terms of simpler ones following a divide-and-conquer strategy \cite{keisuke_deepvqe_2020, zhou_dnc_2022}.

\section{Acknowledgements}
This work was supported by the Leverhulme Trust Research Leadership Award RL-2019-015, by EPSRC grant EP/R020612/1 and by the German Federal Ministry of Education and Research (BMBF) through the funded project
EQUAHUMO (Grant No. 13N16066) within the funding
program quantum technologies - from basic research to
market, in association to the Munich Quantum Valley.
CNS acknowledges financial support from the UK Hub in Quantum Computing and Simulation, part of the UK National Quantum Technologies Programme with funding from UKRI EPSRC grant EP/T001062/1.
KM acknowledges financial support from the Royal Commission for the Exhibition of 1851.
Statement of compliance with EPSRC policy framework on research data: This publication is theoretical work that does not require supporting research data.

\appendix

\renewcommand*{\thesection}{%
  \Alph{section}\texorpdfstring{}{. }%
}
\renewcommand*{\thesubsection}{%
  \Alph{section}.\arabic{subsection}\texorpdfstring{}{. }%
}

\section{Variance of gradient} \label{sec:gradient}

Here, we study how the variance of the gradient with respect to the center angle scales with the lattice size and the circuit depth.  This variance is known to capture the phenomenon of barren plateaus, as it provides an upper bound for the magnitude of the gradients across the optimization landscape \cite{mcclean_barren_2018, holmes_expressibility_2022}. The scaling with respect to the center angle is taken to be representative of the scaling with respect to the other angles \cite{arrasmith_gorges_2021}.

\begin{figure}[hbt]
	\centering
	\includegraphics[width=\linewidth]{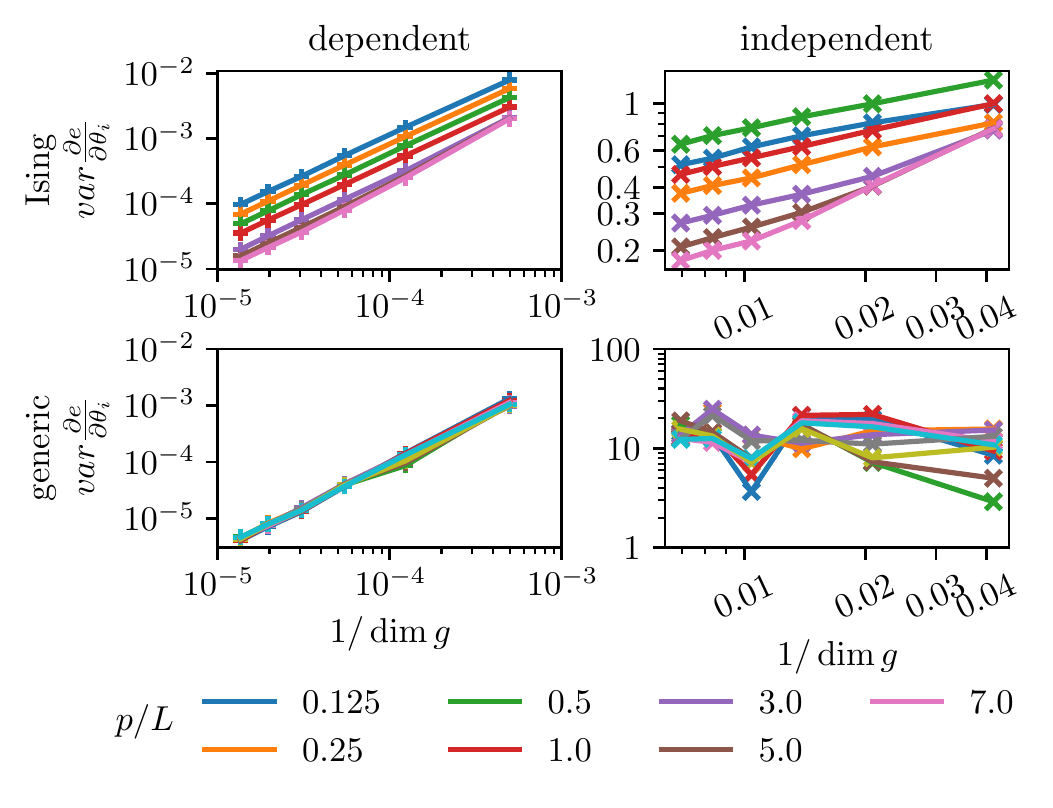}
	\caption{Variance of gradient taken at center angle $|| \partial e/\partial \theta_i ||$ using the site-dependent protocol (left) and the site-independent protocol (right) with the ground state of the Ising model as a target state (top) and $5$ generic quadratic Hamiltonians (symmetric quadratic Hamiltonians in the independent case) as a target state (bottom), plotted against the inverse of the dimension of the Lie algebra generated by the Hamiltonians used in each protocol at different system sizes.  Various values of $p/L$ (bottom labels) between $1$ and $7$ were used. 20000 samples were taken per value of $p$, random state and lattice size.}
	\label{fig:gradient_variance_with_dimg}
\end{figure}

Figure~\ref{fig:gradient_variance_with_dimg} illustrates the scaling of this variance with the lattice size, expressed in terms of the dimension of the Lie algebra, following Ref.~\cite{larocca_control_2021} which conjectured that the variance of the gradient is inversely proportional to this dimension. We see that while this seems to hold in general, it depends on the circuit depth and the state under peparation. When preparing the Ising model, increasing the circuit depth changes this proportionality by a constant factor; while, curiously, when preparing generic FGS with the site dependent protocol, it is independent of the circuit depth. Note also that when the circuit enters the overparameterized regime, for instance when targeting the Ising model in the site-independent case at $p=7L$, this relation can break down as the variance of the gradient saturates at high circuit depths (see next paragraph). Finally, we note that there are exceptions to this relation; in particular, we note that when preparing a generic FGS using the site independent protocol, the gradient seems to oscillate around a constant value of $\sim 10$ without decaying as the system size increases.

\begin{figure}[hbt]
	\centering
	\includegraphics[width=\linewidth]{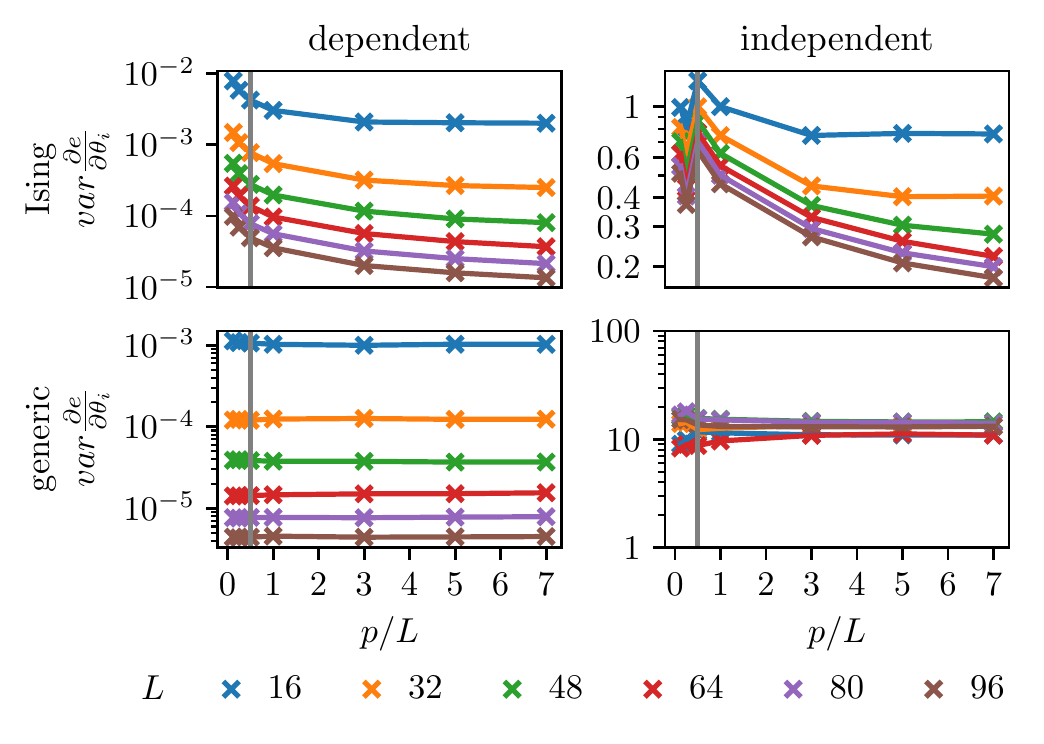}
	\caption{Variance of gradient taken at center angle $||{\partial e}/{\partial \theta_i} ||$ using the site-dependent protocol (left) and the site-independent protocol (right)  with the ground state of the Ising model as a target state (top) and $5$ generic quadratic Hamiltonians (symmetryc generic quadratic Hamiltonians in the independent case) as a target state (bottom), plotted against various values of $p$ between $1$ and $4L$. Several system sizes (bottom labels) were used. 20000 samples were taken per value of $p$, random state and lattice size. The vertical line indicates $p=L/2$.}
	\label{fig:gradient_variance_with_p}
\end{figure}

Figure~\ref{fig:gradient_variance_with_p} illustrates the scaling of the variance of the gradient with circuit depth. We see that when targeting the Ising model, there is an initial drop in this variance, which then stabilizes to a fixed value. In contrast, when preparing a generic FGS, the variance almost immediately converges to this stable value, particularly in the site-dependent case.

\section{Lie algebra structure} \label{sec:lie}

Here, we deduce the structure of the Lie algebras associated with the protocols in Eqs.~\eqref{eq:site_independent_protocol}-\eqref{eq:site_dependent_protocol} for both OBCs and PBCs. We do so by deriving a basis for the algebras generated by the aforementioned sets when unrestricted to any symmetry sector. Then, we restrict these bases to a fixed parity symmetry sector, and see how this affects the structure and dimensions of the algebra. Note that, for notational simplicity, throughout this section we often disregard signs when converting from spin to Majorana operators when it does not influence the generated Lie algebras.

Let us recall the fermionic parity operator
\begin{align*}
  P = \prod_j Z_j.
\end{align*}

\begin{lemma} \label{lem:site_dependent_algebra}
  The Lie algebra generated by $\mathcal{D} = \{iZ_j, iX_j X_{j+1}\}_{j=1,...L}$
  \begin{enumerate}
  \item with OBC is
  \begin{align} \label{eq:dep_algebra_obc}
    \mathfrak{d}_{\text{OBC}} = \{iZ_j, 
    &iX_j Z ... Z X_k, X_j Z ... Z Y_k, \\
    &iY_j Z ... Z X_k, Y_j Z ... Z Y_k: \nonumber \\
    &1 \leq j < k \leq L\} \nonumber \\
    = \{\gamma_j\gamma_k &: 1 \leq j < k \leq L\} &
  \end{align}
  and has dimension $L(2L-1)$.
  \item with PBC is
  \begin{align} \label{eq:dep_algebra_pbc}
    \mathfrak{d}_{\text{PBC}} &= \mathfrak{d}_{\text{OBC}} \cup (P \cdot \mathfrak{d}_{\text{OBC}})
  \end{align}
  where 
  \begin{align}
    (P \cdot \mathfrak{d}_{\text{OBC}}) =  \{
    &iZ ... Z_{j-1} Z_{j+1} ... Z, \quad iZ ... Z X_j X_k Z ... Z, \nonumber \\
    &iZ ... Z X_j Y_k Z ... Z, \quad Z ... Z Y_j X_k Z ... Z, \nonumber \\
    &iZ ... Z Y_j Y_k Z ... Z \quad : 1 \leq j < k \leq L\}
  \end{align}
  and this algebra has dimension $2L(2L - 1)$.
  \end{enumerate}
\end{lemma}

\begin{lemma}
  The Lie algebra generated by $\mathcal{I} = \{i\sum_j Z_j, i\sum_j X_j X_{j+1}\}$
  
    \begin{enumerate}
      \item with OBC is
    \begin{align} \label{eq:indep_algebra_obc}
      \mathfrak{i}_{\text{OBC}} = \{&iZ_j + iZ_{L-j+1}, \\
      &iX_j Z ... Z X_k + iX_{L-k+1} Z ... Z X_{L-j+1}, \nonumber \\
      &iX_j Z ... Z Y_k + iY_{L-k+1} Z ... Z X_{L-j+1}, \nonumber \\
      &iY_j Z ... Z X_k + iX_{L-k+1} Z ... Z Y_{L-j+1}, \nonumber \\
      &iY_j Z ... Z Y_k + iY_{L-k+1} Z ... Z Y_{L-j+1}, \nonumber \\
      &1 \leq j < k \leq \lceil L/2 \rceil \} \nonumber \\
      = \{&\gamma_{2j-1} \gamma_{2j} + \gamma_{2L-2j+1} \gamma_{2L-2j+2},  \\
      &\gamma_{2j} \gamma_{2k-1} + \gamma_{2L-2k+2} \gamma_{2L-2j+1}, \nonumber \\
      &\gamma_{2j} \gamma_{2k} - \gamma_{2L-2k+2} \gamma_{2L-2j+2}, \nonumber \\
      &\gamma_{2j-1} \gamma_{2k-1} - \gamma_{2L-2k+1} \gamma_{2L-2j+1}, \nonumber \\
      &\gamma_{2j-1} \gamma_{2k} + \gamma_{2L-2k+1} \gamma_{2L-2j+2}, \nonumber \\
      &: 1 \leq j < k \leq \lceil L/2 \rceil \} \nonumber,
    \end{align}
    and this algebra has dimension $L^2$ \cite{larocca_control_2021}.

      \item with PBC is
      \begin{align} \label{eq:indep_algebra_pbc}
        \mathfrak{i}_{\text{PBC}} = \{&i\sum_j (Z_j + Z_1 ... Z_{j-1} Z_{j+1} ... Z_L), \\
          &i\sum_j (X_j Z ... Z X_{j+k}  \nonumber \\ 
          &+ Z_1 ... Z_{j-1} X_j X_{j+(L-k)} Z_{j+(L-k)+1} ... Z_L),\nonumber  \\
          &i\sum_j (X_j Z ... Z Y_{j+k} + Y_{j} Z ... Z X_{j+k}\nonumber  \\ 
          &+ Z_1 ... Z_{j-1} X_j Y_{j+(L-k)} Z_{j+(L-k)+1} ... Z_L \nonumber  \\
          &+ Z_1 ... Z_{j-1} Y_j X_{j+(L-k)} Z_{j+(L-k)+1} ... Z_L ), \nonumber  \\
          &i\sum_j (Y_j Z ... Z Y_{j+k} \nonumber \\ 
          & + Z_1 ... Z_{j-1} Y_j Y_{j+(L-k)} Z_{j+(L-k)+1} ... Z_L) \nonumber\\
          & : 1 \leq k \leq L-1\} \nonumber
      \end{align}
    \end{enumerate}
and has dimension $3L-2$.
\end{lemma}

The proofs for these statements follow by inductively taking the brackets of the generators of these algebras. Here, we provide a proof for the structure of $\mathfrak{d}_{\text{OBC}}$; the others are derived in a similar fashion.
\begin{proof}
  By induction on $L$:  
  
  \underline{$L=2$}: Taking the Lie brackets iteratively of $\{iZ_1, Z_2, iX_1 X_2\}$, one obtains the linearly independent set $\{iZ_1, iZ_2, iX_1 X_2, iX_1 Y_2, iY_1 X_2, iY_1 Y_2\}$, which has $6$ elements.
  
  \underline{$L \implies L+1$}: Assume the Lemma holds for $L$. Then, define
  \begin{align}
    \mathcal{G}_{a,b} := \{iZ_j, iX_j X_{j+1}\}_{j=a,...,b}
  \end{align}
  \begin{align}
    \mathcal{L}_{a,b} := \{&iZ_j, iT_{j,k}: a \leq j < k \leq b\}
  \end{align}
  \begin{align}
    T_{j,k} := &A_j \otimes \left (\bigotimes_{m=j+1}^{k-1} Z_m \right ) \otimes B_k, \nonumber \\
    &A_m, B_m \in \{X_m,Y_m\}
  \end{align}
  and let $\mathcal{R}_{a,b}$ be the Lie algebra generated by $\mathcal{G}_{a,b}$. We must prove that $\mathcal{R}_{1,L+1} = \mathcal{L}_{1,L+1}$. 
  
  By induction hypothesis $\mathcal{R}_{1,L}= \mathcal{L}_{1,L}$. Using this, and the definition of Lie algebra generators, we obtain $\mathcal{L}_{1,L} \subseteq \mathcal{R}_{1,L+1}$. Since it is easy to prove that 
  \begin{align}
    \left [iT_{i,j},  iT_{k,l} \right] \propto \delta_{jk} i T_{i,l},
  \end{align}
  and $\left [iT_{i,j},  iZ_{k} \right] \propto (\delta_{ik} + \delta_{jk}) i T_{i,j}$, we conclude that $\mathcal{R}_{1,L+1} \setminus \mathcal{L}_{1,L} \subseteq \{iT_{k,L+1}\}_{k=1,...,L} \cup \{iZ_{L+1}\}$. But $T_{k,L+1} \subset \mathcal{L}_{k, L+1} = \mathcal{R}_{k, L+1} \subseteq \mathcal{R}_{1, L+1}$, and $Z_{L+1} \in \mathcal{G}_{1, L+1}$. Thus it must be the case that $\mathcal{R}_{1,L+1} = \mathcal{L}_{1,L+1}$.
  
  Since, from the above, $\mathcal{L}_{1,L+1} = \mathcal{L}_{1,L} \cup \{T_{k,L+1}\}_{k=1,...,L} \cup \{Z_{L+1}\}$, and since these sets are disjoint, using the induction hypothesis, the dimension of $\mathcal{L}_{1,L+1}$ is $L(2L-1) + 4L + 1 = (L+1)(2(L+1)-1)$
  \end{proof}

Noting that every element of the generators of the Lie algebras commute with $P$, we now state the structure of these algebras restricted to each of the parity symmetry sectors.

We first note that, as can be seen in Lemma~\ref{lem:site_dependent_algebra} for the generators $\mathcal{D}$, when restricting to a parity sector, the algebra in the OBC case remains unchanged, while the algebra in the PBC case is cut in half, and is equal to former. Hence:
\begin{align} \label{eq:parity_dep_algebra}
  \mathfrak{d} :=& \,\mathfrak{d}_{\text{OBC}} = \mathfrak{d}_{\text{OBC}} \Big|_{P = \pm 1} = \mathfrak{d}_{\text{PBC}} \Big|_{P = \pm 1} 
\end{align}
and it has dimension $L(2L-1)$.

As for the case of the set of generators $\mathcal{I}$, we see that the algebra with OBC remains unchanged when restricted to a parity sector. Hence
\begin{align} \label{eq:parity_indep_algebra_obc}
  \mathfrak{i}_{\text{OBC}} = \mathfrak{i}_{\text{OBC}} \Big|_{P = \pm 1}
\end{align}
and it has dimension $L^2$.

Finally, the same set of generators with PBC yields
\begin{align}\label{eq:parity_indep_algebra_pbc}
  \mathfrak{i}_{\text{PBC}}\Big|_{P = \pm 1} = \Bigl\{&i\sum_j Z_j, \\
    &i\sum_{j} (X_j Z ... Z X_{j+k}  \mp Y_j Z ... Z Y_{j+ N - k}), \nonumber \\
    &i\sum_{j} (X_j Z ... Z Y_{j+k} + Y_j Z ... Z X_{j+k} \nonumber \\
    &\pm X_j Z ... Z Y_{j+N-k} \pm Y_j Z ... Z X_{j+N-k}), \nonumber \\
    &i\sum_{j} (Y_j Z ... Z Y_{j+k} \mp X_j Z ... Z X_{j+N-k}) : \nonumber \\
    &1 \leq k \leq L-1 \Bigr\}  \nonumber \\
  = \Bigl\{&\sum_j \gamma_{2j-1} \gamma_{2j}, \nonumber \\
    &\sum_j \gamma_{2j} \gamma_{2(j+k)-1}  \pm \gamma_{2j - 1} \gamma_{2(j + N - k)}, \nonumber \\
    &\sum_j \gamma_{2j} \gamma_{2(j+k)} - \gamma_{2j} \gamma_{2(j+k)-1} \nonumber \\
    &\pm \gamma_{2j} \gamma_{2(j+N-k)-1} \mp \gamma_{2j-1} \gamma_{2(j+N-k)-1}, \nonumber \\
    &\sum_i \gamma_{2j-1} \gamma_{2(j+k)} \pm \gamma_{2j} \gamma_{2(j+N-k)-1} \nonumber \\
    &: 1 \leq k \leq L-1 \Bigr\} 
  \end{align}
and it has dimension $\lfloor 3L/2 \rfloor$.

\section{Mechanism behind staircases} \label{sec:overlaps}

\begin{figure}[hbt]
	\centering
	\includegraphics[width=\linewidth]{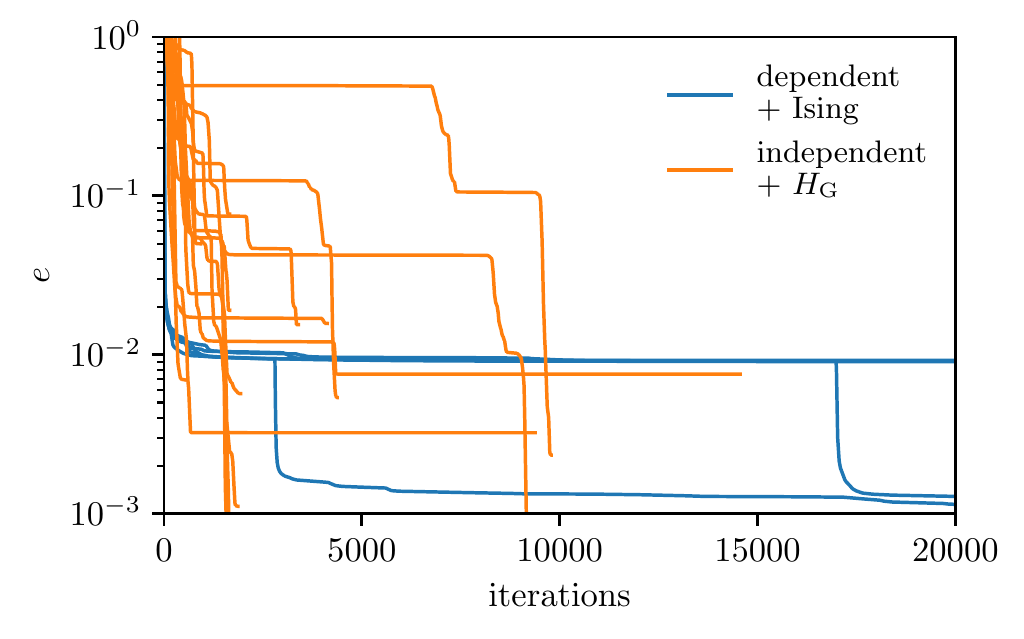}
	\caption{Optimization traces showing  the ``staircase" pattern previously seen in Figure~\ref{fig:iter_vs_cost}. Here, the system size is $40$, and the label refers to the protocol used and the Hamiltonian targetted, respectively. The target Hamiltonian is either the Ising model, Eq.~\eqref{eq:ising}, or a generic symmetric quadratic Hamiltonian, Eq.~\eqref{eq:random}). PBCs and exact parameterization ($p=L/2$) are assumed.}
	\label{fig:staircase}
\end{figure}

In Section~\ref{sec:traces}, we described a pattern that  we dubbed "staircase", where the optimizer gets stuck and the value of the cost function undergoes very little variation for a number of iterations until it sharply drops to a new plateau; we highlight  this phenomenon more clearly in Figure~\ref{fig:staircase}. Here, we offer an explanation for this observation.

\begin{figure}[hbt]
	\centering
	\includegraphics[width=\linewidth]{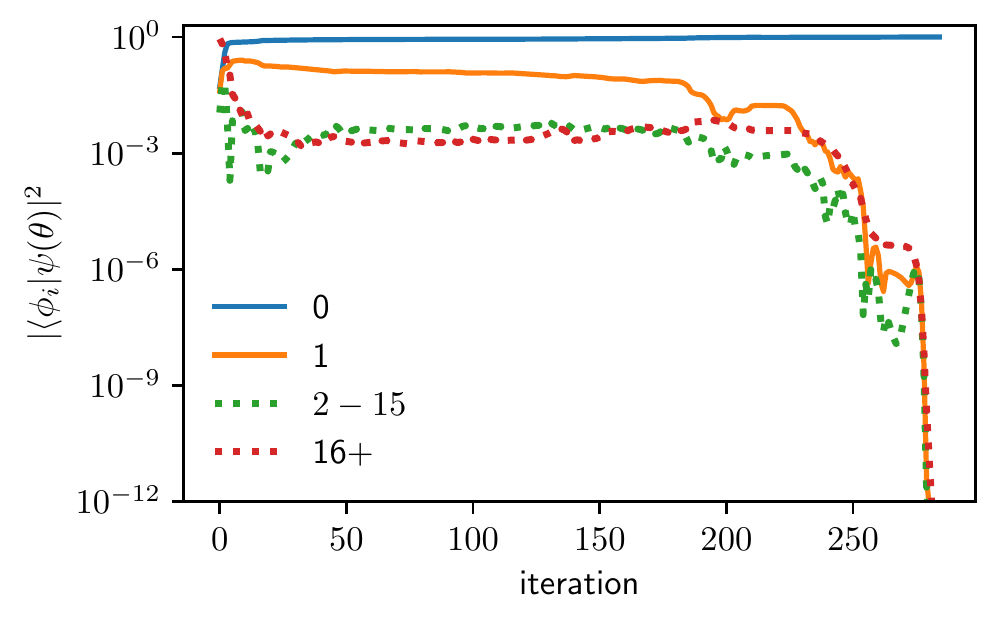}
	\caption{Overlaps of lowest excited states of the target Hamiltonian with the state under preparation $\ket{\psi(\theta_i)}$. Legend indicates the index of an eigenstate with which overlaps were taken, with $0$ indicating the ground state, $1$ indicating the first excited state, etc., while the ranges indicate a sum of the overlaps with the corresponding eigenstates. We see a high overlap with the first excited state throughout the optimization, indicating that the cost function cannot easily distinguish this state from the ground state. The site-independent protocol was used and the Ising model \eqref{eq:ising} was targeted at the exactly parameterized depth $p=L/2$ for $L=16$.}
	\label{fig:overlaps}
\end{figure}

In Figure~\ref{fig:overlaps}, we plot the overlap of the state along the optimization with the eigenstates of the target Hamiltonian. We notice that the overlap of the state under preparation with the target Hamiltonian is orders of magnitude higher than with other excited eigenstates. The dynamics of state preparation is thus dominated by a competition between the ground state and the first excited state. We propose that the staircase plateaus we observe follow a similar mechanism: in each plateau, there is a state in the Hilbert space (akin to the first excited state in the previous description) that fully captures the features that the cost function struggles to distinguish from those of the ground state in each of these plateaus.

\section{Scaling of Hessian and optimization} \label{sec:hessian}

\begin{figure*}[hbt]
	\centering
	\includegraphics[width=\linewidth]{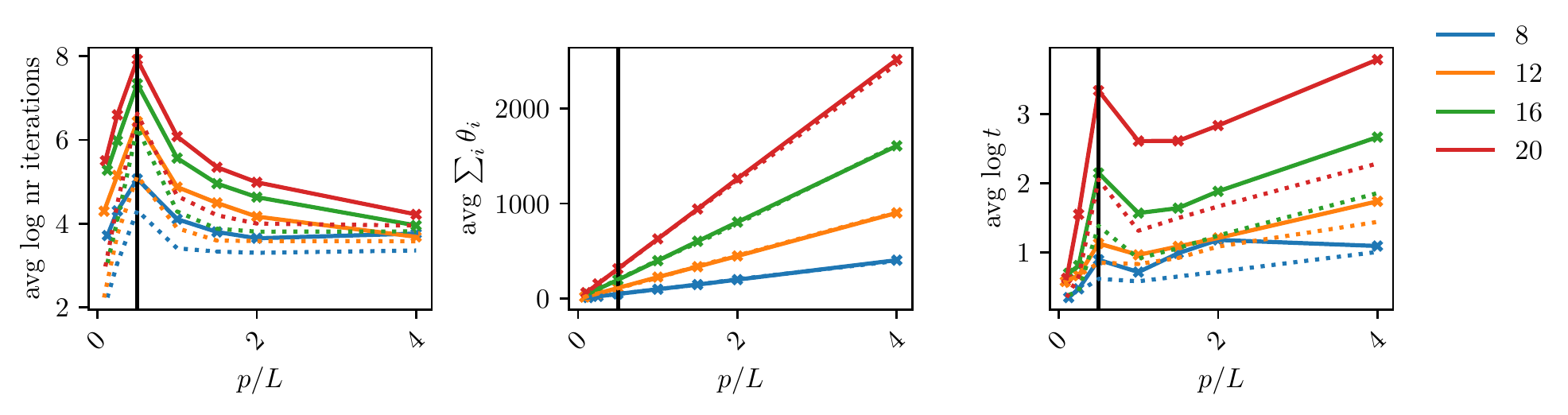}
	\caption{Different quantities characterizing the hardness of the optimization with increasing circuit depth. On the left, we plot the mean of the logarithm of the number iterations to converge; the center plot depicts the mean of the sum of all the angles of the protocol, where periodicity is appropriately taken into account; on the right, the average of the logarithm of the total computational time is shown.  Generic symmetric quadratic Hamiltonians were targeted, and results were averaged over 5 random states and 5 random initializations per state. Filled line corresponds to the site-dependent protocol, while dotted line represents the site-independent protocol; these two cases essentially display the same behavior. Periodic boundary conditions were used, and the black vertical line indicates $p=L/2$, the depth at which the circuit is exactly parameterized. These results were obtained on an Intel(R) Xeon(R) CPU E5-2650 v4 @ 2.20GHz.}
	\label{fig:hessian_memory}
\end{figure*}

In Figure~\ref{fig:hessian_memory}, we examine the effect of increasing the circuit depth on the total time taken to run an optimization and on the time it would take to prepare these states on a quantum simulator (measured by the sum of the angles). We see that despite increasing the circuit depth into the overparameterized regime making the optimization easier, the sum of the angles in the protocol grows linearly with the circuit depth, increasing the time to run the circuit on a quantum simulator and making it more prone to errors. Furthermore, we see that despite the number of iterations to converge decreasing into the overparameterized regime, when one looks at the actual time taken to run the optimizer, there is an inflexion point where it first goes down and then starts increasing again. This is due to the algorithm (BFGS) used, which stores an approximation to the Hessian; as the size of the Hessian increases, the computational cost associated to storing and manipulating it dominates the computational time. Thus, while it is feasible to a larger class of optimization algorithms at lower system sizes, as one increases the circuit depth, one has to switch to algorithms specialized to dealing with a large number of parameters e.g. ADAM or stochastic gradient descent.

\bibliographystyle{plainnat}
\bibliography{bibliography}
\end{document}